\documentclass[twocolumn]{aastex63}
\usepackage{rotating}

\newcommand\mkp{\texttt{mkidpipeline} }

\received{January 6, 2022}
\revised{February 21, 2022}
\accepted{February 22. 2022}
\submitjournal{AJ}

\shorttitle{MKID Pipeline}
\shortauthors{Steiger, Bailey et al.}
\graphicspath{{./}{figures/}}

\begin{document}

\title{The MKID Pipeline: A Data Reduction and Analysis Pipeline for UVOIR MKID Data}

\author[0000-0002-4787-3285]{Sarah Steiger}
\thanks{These authors contributed equally}
\affiliation{Department of Physics, University of California, Santa Barbara, Santa Barbara, California, USA}

\author[0000-0002-4272-263X]{John I. Bailey, III}
\thanks{These authors contributed equally}
\correspondingauthor{John I. Bailey, III}
\email{baileyji@ucsb.edu}
\affiliation{Department of Physics, University of California, Santa Barbara, Santa Barbara, California, USA}

\author[0000-0003-3146-7263]{Nicholas Zobrist}
\affiliation{Department of Physics, University of California, Santa Barbara, Santa Barbara, California, USA}

\author[0000-0001-5721-8973]{Noah Swimmer}
\affiliation{Department of Physics, University of California, Santa Barbara, Santa Barbara, California, USA}
\author{Rupert Dodkins}
\affiliation{Department of Physics, University of California, Santa Barbara, Santa Barbara, California, USA}

\author[0000-0001-5587-845X]{Kristina K. Davis}
\affiliation{Department of Physics, University of California, Santa Barbara, Santa Barbara, California, USA}

\author[0000-0003-0526-1114]{Benjamin A. Mazin}
\affiliation{Department of Physics, University of California, Santa Barbara, Santa Barbara, California, USA}



\begin{abstract}

We present The MKID Pipeline, a general use science data pipeline for the reduction and analysis of ultraviolet, optical and infrared (UVOIR) Microwave Kinetic Inductance Detector (MKID) data sets. This paper provides an introduction to the nature of MKID data sets, an overview of the calibration steps included in the pipeline, and an introduction to the implementation of the software. 

\end{abstract}


\section{Introduction} \label{sec:intro}


Microwave Kinetic Inductance Detectors (MKIDs) are photon detectors comprised of an array of RF-multiplexed inductor-capacitor resonators capable of measuring both the arrival time and wavelength ($R\sim 4-40$) of individual photons without read noise or dark current \citep{day2003broadband, Szypryt2017}. 
To date,
MKID-based instruments 
have placed new constraints on pulsar timing \citep{strader2016search, collura2017search} and the orbital decay of compact binaries \citep{szypryt2014direct}. Additionally, their ability to discriminate photons at the $10^{-5}$~s level is crucial to improving contrast limits for exoplanet direct imaging: it enables real-time starlight suppression via an extreme adaptive optics feedback loop \citep{fruitwala2021readout} and separation of faint companions from  diffracted starlight via the statistical analysis of photon arrival times with stotochastic speckle discrimination \citep[][]{meeker2018darkness, walter2019stochastic, steiger2021}.

Each pixel in an MKID detector is a lithographed lumped element superconducting inductor-capacitor resonator circuit. When a photon is absorbed by the inductor,
the energy breaks Cooper pairs causing a change of inductance that can be measured as a change in the resonator's phase. Room temperature readout electronics monitor each resonator's phase with 1 MHz frequency \citep{Fruitwala2020} yielding microsecond timing precision. Since the amplitude of this phase change is proportional to the number of Cooper pairs broken, the energy (wavelength) of the incident photon can also be determined to within 5-10\%. See \citet{mazin2012} and \citet{Szypryt2017} for more information. 

MKIDs produce a raw data stream that differs from typical semiconductor based astronomical detectors and requires significant post-processing before it is effectively accessible to the broader astronomical community. To this end we have created the MKID Pipeline\footnote{https://github.com/MazinLab/MKIDPipeline}, a Python package to provide an open-source extensible data reduction pipeline for MKID data. This pipeline takes raw MKID data as input and processes it into either a traditional form (i.e. FITS cubes) to be used with existing astronomical analysis packages, or a unique MKID data product suitable for advanced analysis tailored to the detector's unique abilities.

The MKID Pipeline is based on the development and use of the only three optical/near-infrared astronomical MKID instruments to date: ARCONS \citep{mazin2013arcons}, DARKNESS \citep{meeker2018darkness}, and the MKID Exoplanet Camera \citep[MEC,][]{walter2020}. MEC is a new user instrument located behind the Subaru Conoragraphic Extreme Adaptive Optics System (SCExAO) at the Subaru Telescope on Maunakea \citep{Jovanovic2015}. Though largely targeted to MEC and our current laboratory development, the pipeline is intended to be easily extensible to future instruments.

In this paper, we first briefly describe the contents of a typical MKID observing data set (\S \ref{sec:mkidobserving}). We then discuss how data is processed in \S \ref{sec:dataprocessing}, beginning with a description of the contents of raw MKID data before diving into the specific calibration algorithms in depth. Finally, in \S \ref{sec:implementation} we end with a discussion of how these steps are implemented in software and how a user would perform basic data reduction. 

\section{MKID Observing Data Sets}
\label{sec:mkidobserving}

MKID detectors take data by recording the time, location, and phase response for each detected photon. For this reason, all time binning is performed in post processing. An MKID observation or `exposure' therefore refers not to specific exposures determined during a night of observing, but to time ranges where the object of interest is on the detector at an intended position. The resulting total observational data set consists of some number of science observations, associated observatory and instrumental metadata, and necessary calibration data. 

Science observations consist of a single time range, target, sky position, and associated calibration data sets. Due to the current high level of detector defects (e.g. cold/dead pixels), it is common to take dithered data suitable for reconstruction of a sky mosaic \citep{2000dither}. In MEC, a tip/tilt mirror is used for this purpose. This dither then consist of a series of science observations and corresponding tip/tilt mirror positions that are combined in post processing to generate a single output image (see \S \ref{sec:drizzler}). 

Calibration data consists of a series of uniform monochromatic laser exposures relatively evenly spaced across the wavelength coverage of the detector.  These exposures are used for wavelength calibration (\S \ref{sec:wavecal}) and can also be used for flat-fielding (\S \ref{sec:flatcal}), though sky flats may also be taken and used instead. Dark observations (intervals obtained with a closed shutter or on blank sky) may be included to remove instrumental or astrophysical backgrounds. Finally, support for observations of an astrometric reference is provided to calibrate the final output products to reflect real on-sky coordinates (\S \ref{sec:wcscal}), though this data is not routinely required.

Both science and calibration data has associated observatory and instrumental metadata (e.g. observatory and telescope status information, detector temperatures). The instrument control software records all of this data periodically in a machine readable format for later use by the pipeline. A subset of this data must be provided either via these logs or specified by the user when defining data to ensure proper reduction.  

\section{Data Processing}\label{sec:dataprocessing}
\begin{sidewaysfigure*}[ht]
\vspace{6cm}
\includegraphics[width=\textwidth]{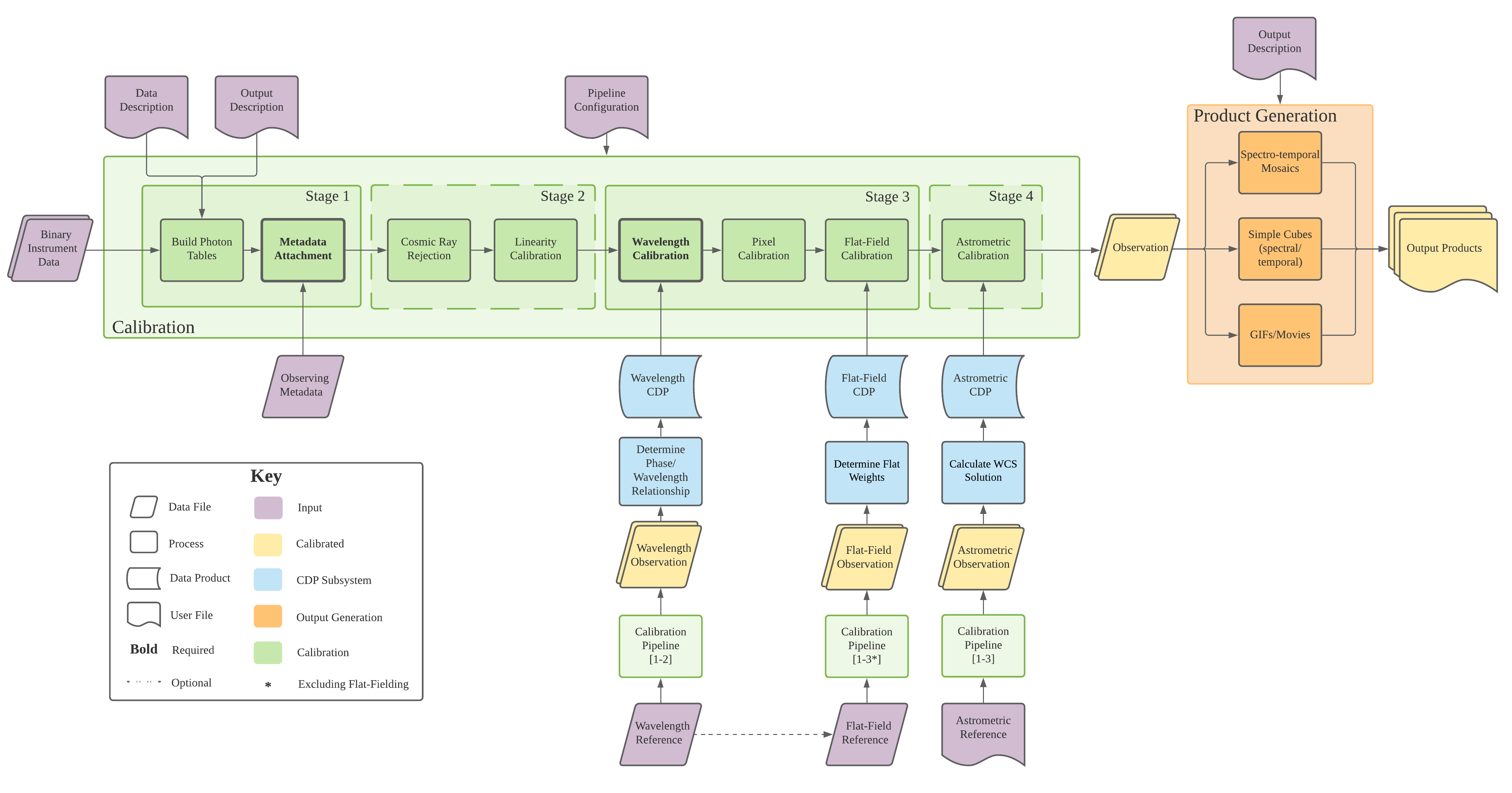}\label{fig:block2}
\caption{A block diagram of the MKID Pipeline.  Processing steps are shown as rectangles with calibration steps in green, product generation steps in orange, and processes that are a part of the calibration data product (CDP) subsystem in blue. Files containing observing data (in various formats) are depicted by purple (input) and yellow (processed) rhombuses. User inputs are in purple. Calibration data products themselves are shown as blue cylinder faces. Rectangles with the wavy bottom edge represent files where the purple are user input configuration files and the yellow are the files representing calibrated output data products. All calibration steps are optional with the exception of the bolded 'Metadata Attachment' and 'Wavelength Calibration' steps without which product generation cannot occur. The dashed stages of the pipeline may be completely omitted. The bracketed numbers in the 'Calibration Pipeline' boxes  denote pipeline stages that must completed to  generate the respective CDPs. Flat-fielding can be accomplished via reference to wavelength calibration data, denoted by the dashed arrow.}
\end{sidewaysfigure*}

Raw MKID data consists of per-resonant-frequency (an analog to pixel) time series of photon-induced phase shifts. These are associated with individual pixels, converted to tabulated photon event data for each observation, and calibrated via the pipeline diagrammed in Figure \ref{fig:block2}. 

In brief, the telescope and instrument logs (along with user overrides) are first used to create an associated metadata time series for for each observation to properly carry out later steps and determine eventual FITS and output header keys. Cosmic rays are then identified within the photon list. A linearity calibration may be performed which calculates a weight for each photon to statistically correct for missing photons caused by a detector-imposed dead time inherent to the MKID readout. This dead time prevents the recording of a photon that arrives too close the the tail of the preceding photon and causes non linear  responses at high count rates ($\gtrsim 5000~photon~pixel^{-1}$, exceeding current instrument limitations).
A series of monochromatic exposures is next used to determine the relationship between phase shift and wavelength for each pixel. Pixels that exhibit too strong (hot), too weak (cold), or no (dead) response to incoming photons are then masked and ignored in further analysis. Inter-pixel variations are next corrected by using an uniform polychromatic exposure or set of monochromatic exposures to determine a spectrally dependent flat weight for each pixel. Finally, an astrometric reference can be used to determine the pixel to world coordinate system (WCS) mapping for the instrument to yield physical output units for both the spatial and spectral dimensions of the output.

The resulting calibrated data is then used to create output products such as spectro-temporal FITS cubes, calibrated tables of photons, and movies. This section describes the algorithmic details of each calibration step outlined above. For details on the implementation of the pipeline itself, see \S \ref{sec:implementation}.

\subsection{Data Format}
MKID detectors are read out via frequency multiplexing sets of pixels that share a microwave feedline. Photon arrival locations are therefore discriminated by frequency rather than detector position. This means the resulting raw MKID data is a per-resonant-frequency time-series of photon-induced phase shifts. Due to the potential for data rates up to $40~MB~s^{-1}~kpix^{-1}$ the data is recorded in a packed binary format  \citep{Fruitwala2020}. This, coupled with the environmental sensitivity of MKIDs, necessitates the occasional determination of a optical beam position to pixel frequency mapping. At the start of pipeline processing this mapping, or ``beammap''---which also contains information about malformed, inoperative pixels---is used to ingest this packed binary data and produce a tabulated photon list for further processing. 

\subsection{Data Calibration}
\subsubsection{Metadata Attachment}

During observing, the instrument captures a record of telescope and instrument status information in addition to the photon data from the detector. After photon table construction, this data is parsed for records within the observing interval as well as the record immediately prior, forming a metadata time-series for each. These series, supplemented with any user specified values, are attached to the photon table.

\subsubsection{Cosmic Ray Rejection}
Cosmic rays incident on an MKID detector excite phonons in the detector substrate causing the majority of pixels to register photon events near-simultaneously for a brief duration. The cosmic ray rejection step identifies intervals where these false photons are recorded for use in later analysis. This is done by splitting observations into $\sim10$ $\mu$s time bins and using one of two techniques to compute a count rate above which a cosmic ray event is flagged.


\begin{figure}[ht!]
\plotone{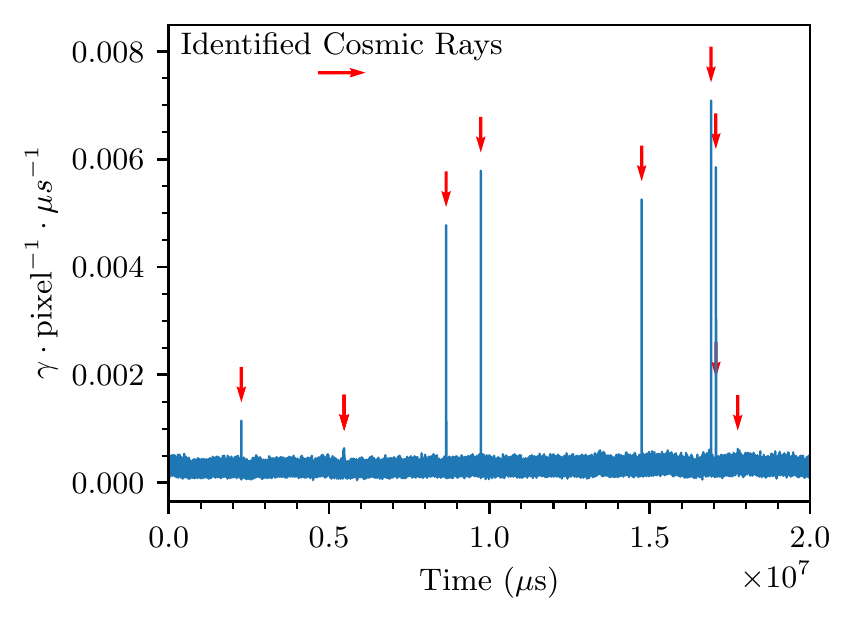}
\caption{Photon time stream where the red arrows denote locations of identified cosmic rays. Excluding all photon data obtained in a 10 $\mu$s window around each event would eliminate a total of 0.00195 s, $< 0.01 \%$ of the exposure. Any missed cosmic rays would contribute no more than a single photon per pixel in an astrophysical source.  
\label{fig:cosmiccal}}
\end{figure}

The first approach assumes that count rates should follow Poisson statistics and employs \texttt{scipy.stats} to generate a count rate threshold \citep{2020SciPy-NMeth}. First, a cumulative density function (CDF) is determined which is defined by the number of standard deviations away from the mean that a given count rate needs to be for that time bin to be classified as containing a cosmic ray. A percent point function is then evaluated on that CDF at the average count rate to generate the threshold value. The second method calculates the standard deviation of the count rates using the total binned time stream, excluding data that falls outside of three standard deviations from the mean. The threshold is then defined as a user input number of those standard deviations above the mean value. 

In both cases, bins that exceed the computed threshold are flagged as cosmic ray events and their time intervals, total and average counts, and peak count rates recorded in the photon table's header. Due to the microsecond timing resolution of MKIDs the total time lost due to cosmic rays in a typical data set is less than $0.01\%$ of the total observation time. In contrast to a CCD detector, missed events would only add a single count to each pixel. For this reason, cosmic ray rejection is presently implemented in a way that does not alter the original photon time stream and removal is not merited unless a particular analysis is sensitive to false counts at the 10s of photons level.  
Figure \ref{fig:cosmiccal} shows an example MKID photon time stream with cosmic rays identified. 

\subsubsection{Wavelength Calibration} \label{sec:wavecal}

\begin{figure*}[ht!]
\centering
\includegraphics[width=\textwidth]{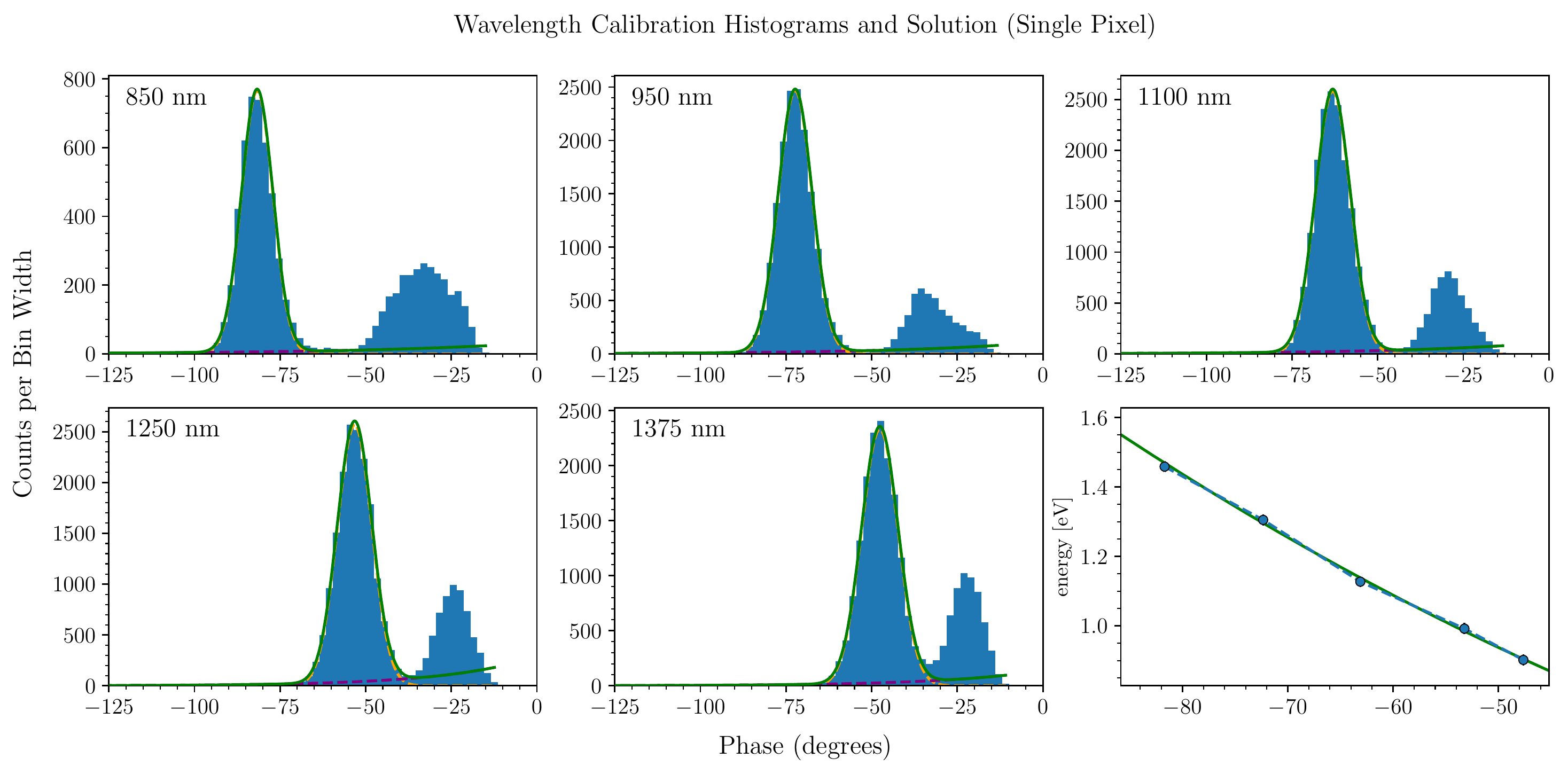}
\caption{Single pixel count rate histograms for each laser wavelength as well as the calibration solution fit (bottom right). The small Gaussian bump at low phases is likely due to an IR leak around 2.7~$\mu$m in the filter stack of the instrument (MEC) used to take this data.\label{fig:wavecal}}
\end{figure*}

The wavelength calibration calculates the relationship between the phase response of each pixel and the wavelength of each incident photon via phase pulse-height histograms generated from a series of monochromatic laser exposures. These exposures are typically generated by using a series of lasers spanning the wavelength sensitivity range of the particular instrument coupled with an integrating sphere to ensure a uniform illumination on the array. 

The phase histograms are fit using one or more of a series of models. Current supported models are a Gaussian signal plus a Gaussian background, and a Gaussian signal plus an exponential background. If more than one model is specified then all are attempted and the best fit one used. When provided, a dark observation is used to subtract a background count rate from the phase histograms to yield a better fit. 

\begin{figure}[ht!]
\plotone{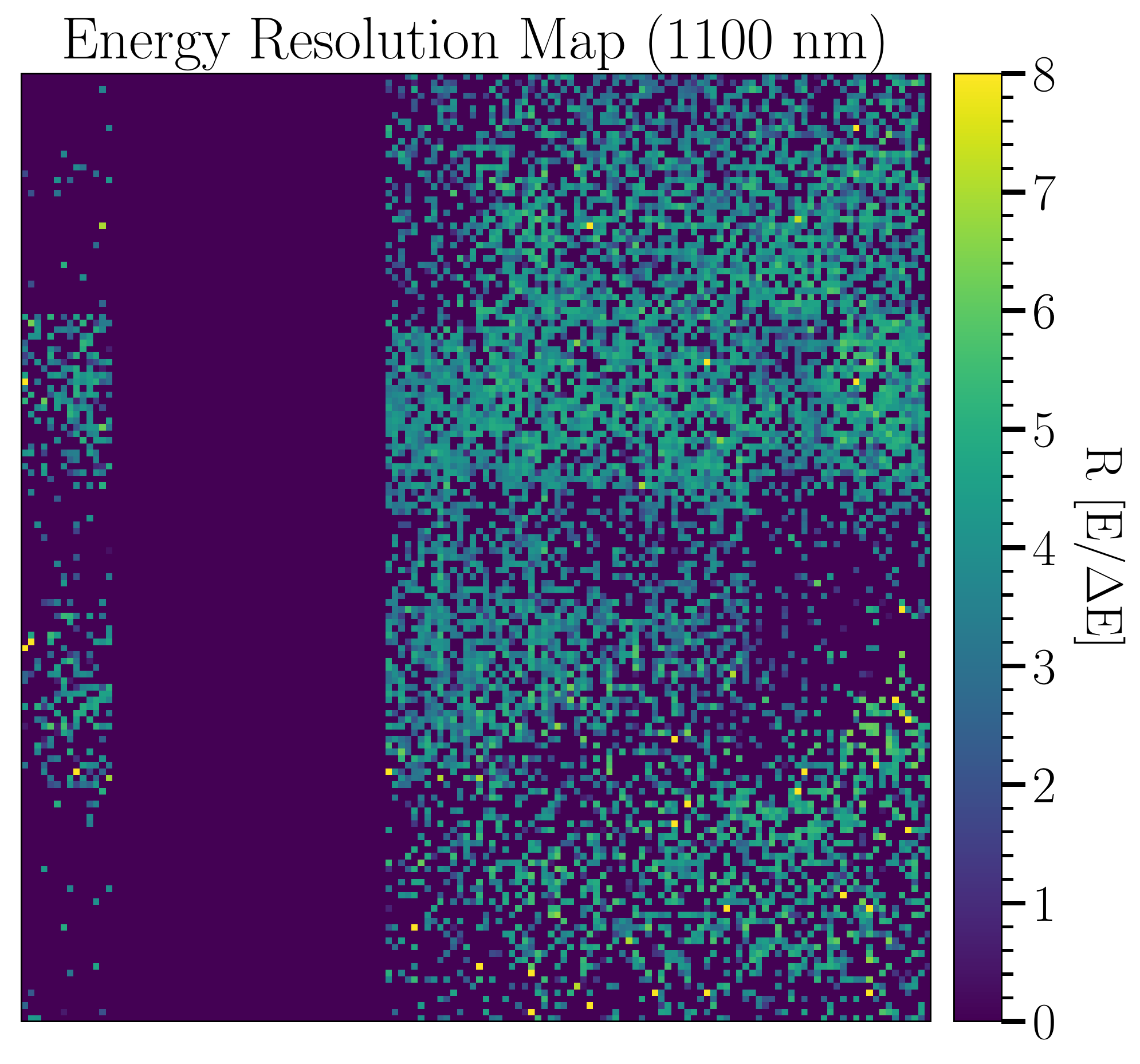}
\caption{Resolution image at 1.1 $\mu m$ for the detector in MEC as of the time of this publication. The median energy resolution (R) across the array at this wavelength is 3.93 excluding dead pixels. This particular detector has three defective feed lines (each containing 2000 pixels) which results in the large strip of dead (R=0) pixels seen to the left of the image \label{fig:array_res}}
\end{figure}

Once the phase histograms are fit, the centers of each histogram are determined and fit as a function of laser wavelength with a linear or quadratic function to determine a final phase-wavelength calibration for each pixel (Fig. \ref{fig:wavecal}). The resulting fits constitute a wavelength calibration data product that consists of a per-pixel mapping of phase to wavelength, a set of associated calibration quality flags, and general solution metadata. A sample resolution map at 1.1 $\mu m$ is shown in Figure \ref{fig:array_res} 

Individual observations are then calibrated using the appropriate (e.g. user-specified, temporally proximate) solution for a given observation by loading each pixels' phases and feeding them through the associated mapping. The resulting wavelengths, associated flags, and wavelength calibration metadata are then stored in the observation. 

\subsubsection{Pixel Calibration}

The pixel calibration identifies `hot', `cold', and `dead' pixels to be removed from further analysis. Pixels that register counts a specified number of intervals above a threshold are flagged hot, below a threshold cold. Dead pixels first determined based on the detector's beammap and the array image is then passed through a filter which iteratively replaces the dead pixels with the mean value of pixels in a surrounding box until none remain. This is done before the determination of hot and cold pixels so as to not skew the algorithms. Three algorithms are provided for determining the hot and cold thresholds and associated interval for each pixel. 

\begin{figure}
\plotone{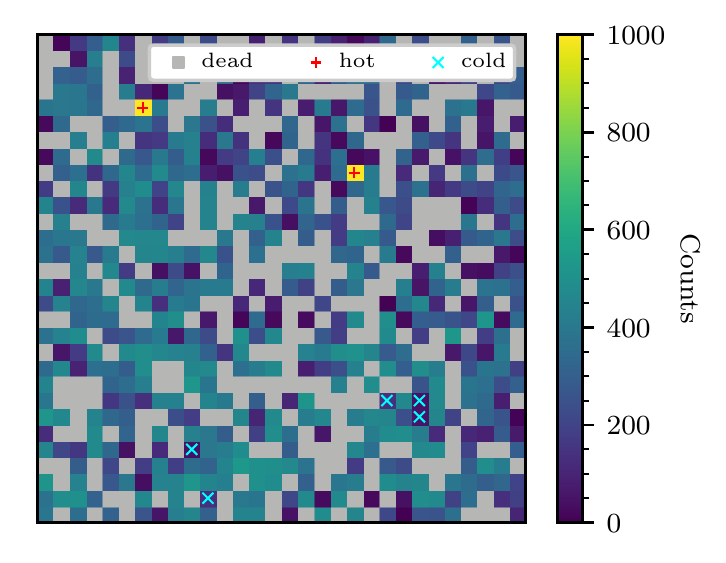}
\caption{Subset of an MKID array with hot, cold, and dead pixels labeled. The threshold method was used in the determination of the pixel flags with default settings. \label{fig:pixcal}}
\end{figure}

\textbf{Threshold}
This method compares the ratio of the image to a moving-box median that excludes both the central pixel and any defective pixels. Ratios greater than some tolerance above/below the peak-to-median of a Gaussian PSF are flagged as hot/cold respectively. See Figure \ref{fig:pixcal} for a sample of this algorithm used on a data set. 
Care must be exercised to ensure the moving box is sufficiently large to not be biased by clusters of hot or cold pixels.

\textbf{Median} The detector array's median count value is used as the global threshold. The tolerance interval is determined by applying a standard deviation moving box filter to the counts image. 

\textbf{Laplacian}
A Laplace filter (\texttt{scipy.ndimage. filters.laplace}) is applied to the image and the result adopted as the count threshold. The standard deviation of the filtered image is used for the tolerance interval. 

\subsubsection{Linearity Calibration} \label{sec:linearity}
Each pixel has a finite dead time, imposed in firmware, that precludes detection of photons arriving within a small time interval following the preceding photon. The exact interval value depends on the quasi-particle recombination time of the superconducting film and the LC time-constant of the resonator. For MEC, this dead time is set in firmware to be $\sim10 \mu s$. As a result, MKID detectors exhibit a nonlinear response that requires correction at high count rates (see Fig. 15 of \citet{arconspipeline}). This correction is equal to $(1 - N \cdot \tau / T)^{-1}$ where N is the number of detected photons in time T for a pixel with dead time $\tau$. The time T is set by the user and should be small so as to effectively determine the instantaneous count rate for each photon. 

The need to compute and operate this calibration on per-pixel inter-photon arrival times can result in expensive computation, especially as single exposures may easily contain $>10^9$ photons. As the effect is less than one part in 1000 for typical count rates, the use of this step is generally discouraged. 

\subsubsection{Flat-field Calibration} \label{sec:flatcal}

Flat-field calibration has two modes: laser and white light. In both modes, a spectro-temporal cube is generated and used to determine the per-pixel wavelength response weights necessary to achieve a uniform response across the detector array. To calculate this weight, the cube is normalized by the integrated average flux at each wavelength and then a user-specified number of the highest and lowest flux temporal bins are excluded to control for time-dependent contamination of the flat, e.g. radio frequency interference. The average of the remaining temporal bins is then fit as a polynomial function of wavelength and the fit saved as the flat-field calibration data product for later application. Data is flat-fielded by evaluating the polynomial at the wavelength of each photon and incorporating the resulting spectral weight into the photon table.

\textbf{White Light Mode} Uses an observation of a uniform continuum source (e.g. twilight, dome) to generate the spectral cube. In this mode, the spectral sampling is determined by the nominal wavelength resolution of the associated wavelength calibration. 

\begin{figure}[ht!]
\plotone{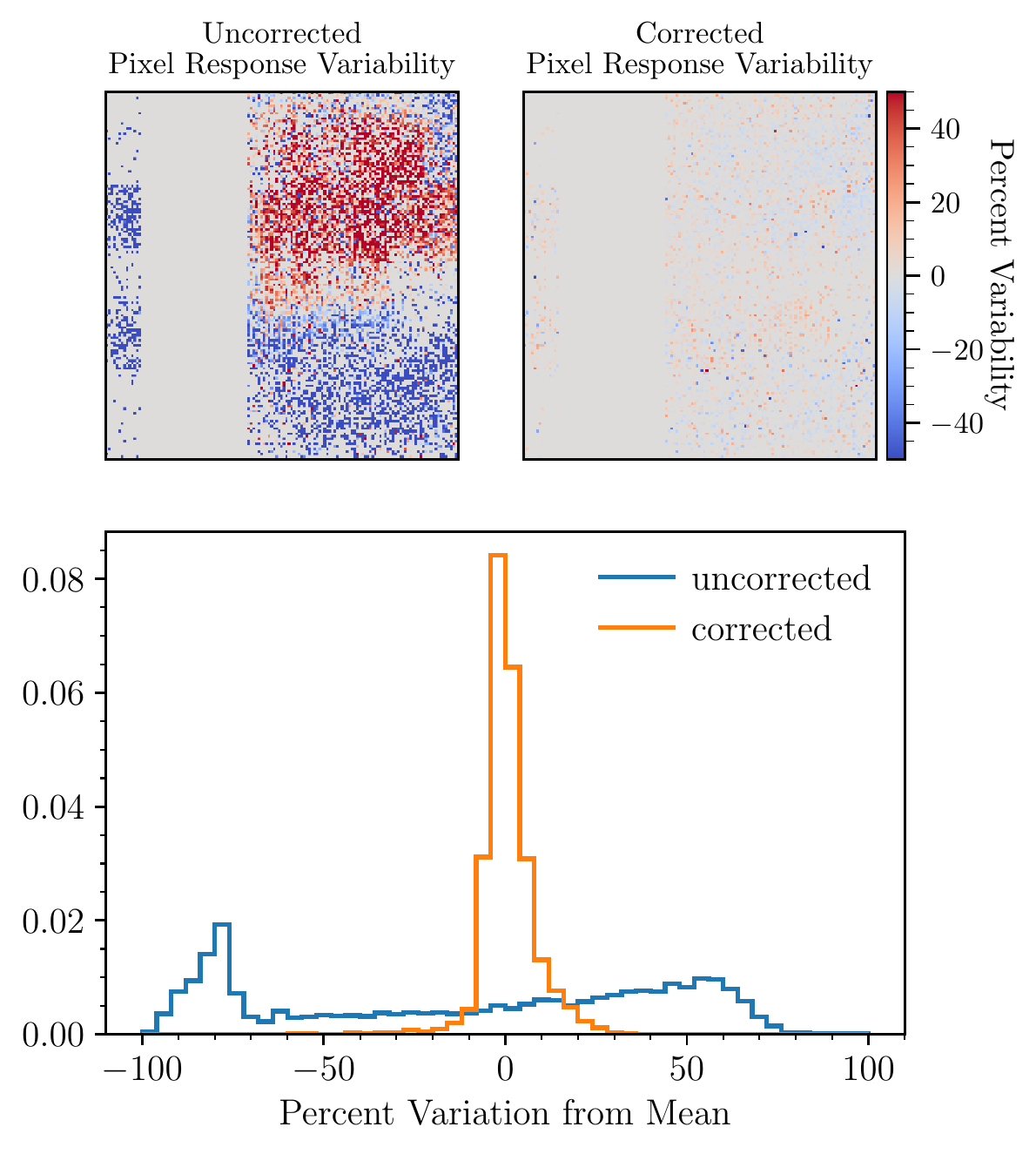}
\caption{Top: Percent variability in pixel response before and after applying the flat-field calibration. This is calculated by subtracting and then dividing the median counts registered on the detector from each pixel. The structure seen in this MEC data is dominated by vignetting from the optical system. Bottom: Histograms of the percent variability with the uncorrected pixel response shown in blue and the corrected pixel response shown in orange. \label{fig:flatcal}}
\end{figure}

\textbf{Laser Mode} Generates the spectral cube using a series of monochromatic laser exposures such as the ones used for the wavelength calibration (see \ref{sec:wavecal}). This can be done by either positing that the laser frames are truly monochromatic (i.e. not imposing any wavelength cut on the exposures), or by using the wavelength calibration solution to use only photons within a small window around each laser wavelength. An example of the flat-field calibration using the laser mode applied to a real data set is shown in Figure \ref{fig:flatcal}.

\subsubsection{Astrometric Calibration}\label{sec:wcscal}

The astrometric calibration determines the World Coordinate System (WCS) transformation parameters to convert an image from pixel (x, y) to on-sky (RA, Dec.) coordinates. First, a point spread function (PSF) fit is performed to determine the pixel location of each source in each image of the observation. Here an `image' is defined by any single exposure where the pixel and sky locations of the sources are expected to remain constant (e.g. the telescope pointing does not change, the tip/tilt mirror is in the same position, etc.).

Each fit PSF location is then assigned an RA and Dec. through the use of an interactive tool where the user selects the approximate pixel location of the PSF for each source coordinate. The fit position of the nearest PSF to the selected coordinates is then assigned to the corresponding sky coordinate to generate a dictionary of pixel-sky coordinate pairs. When complete, the transformation between pixel and sky coordinate is then determined by solving for the WCS parameters by performing the following. 

First, the tip/tilt mirror to pixel mapping is determined by fitting a linear model to the PSF centers ($p_x$, $p_y$) and corresponding mirror positions ($c_x$, $c_y$).

\begin{equation}
    p_{x, y} = \mu_{x, y}c_{x, y} + a_{x, y}
\end{equation}

Here, the slopes $\mu_{x, y}$ give the number of pixels moved for a given tip/tilt mirror position change in either x or y, and $a_{x, y,}$ is the pixel location corresponding to tip/tilt mirror position (0, 0).    

Next, the x and y platescales ($\eta_x$, $\eta_y$) are found using the known separation and pixel displacement of the sources. The platescale is calculated for each image and the mean value saved. 

Finally, an affine transform is applied to the pixel coordinate point consisting of the following steps: 

\begin{enumerate}
    \item Rotation by an angle $\Phi$ to account for the detector's rotation with respect to the telescope beam
    \item Translation by an amount ($\mu_x$$c_x$, $\mu_y$$c_y$) where $c_x$ and $c_y$ are the tip/tilt mirror positions. 
    \item Scaling by the platescale ($\eta_x$, $\eta_y$)
    \item Rotation by the telescope position angle ($\Theta$)
\end{enumerate}



The (RA, Dec.) telescope offset is then added to the transformed pixel coordinate to complete the mapping. This results in two equations for each image ($n_{im}$) and each coordinate pair ($n_{s}$) giving a total of $n_{im} \cdot n_{s} \cdot 2$ equations. Each equation is solved for the last unknown WCS variable, the detector rotation $\Phi$, using \texttt{scipy.optimize.fsolve} and the mean value saved. Values of $\mu_x$, $\mu_y$, $\eta_x$, $\eta_y$, and $\Phi$ are all saved within the photon table metadata.  

\subsection{Data Products}

The calibrated photon tables output by the calibration stage of the pipeline consist of rows of individual photons with columns of time, resonator ID, wavelength and weight. The resonator ID is a unique five to six digit identifying number given to each pixel to determine its location on the array in conjunction with the beammap. The weights are the multiplicative combination of the linearity and flat-field calibration steps. These photon tables may be used directly for analyses that rely on photon arrival time information, such as stochastic speckle discrimination \cite[see][]{fitzgerald2006speckle, gladysz2008, steiger2021}. 

The pipeline is also able to produce traditional astronomical outputs in the form of spectro-temporal cubes from individual observations or dithered mosaics and movies as are described below. Spectral and temporal FITS cubes with arbitrary wavelength and time bin widths may also be generated from individual exposures.

\subsubsection{Spectro-temporal Mosaics}\label{sec:drizzler}

\begin{figure*}[ht!]
\plotone{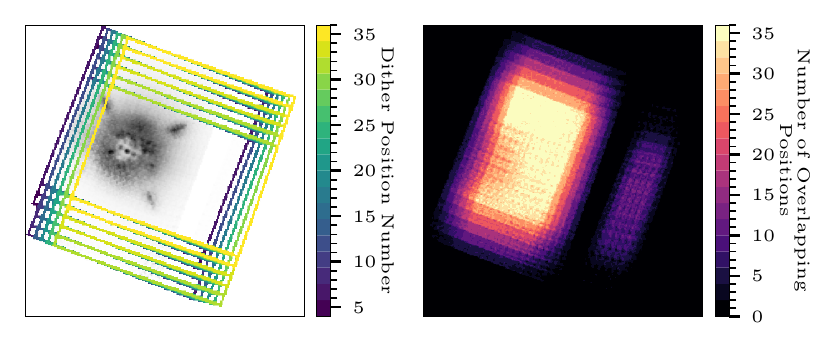}
\caption{Left: MEC Image of the HIP 109427 system with each dither position color-coded by its order in the sequence. The image of the HIP 109427 (behind a coronagraph), satellite spots, and stellar companion are shown in grey scale to be able to better see the frame boundaries. Right: Exposure coverage footprint for the same data set. Here bright regions have more effective exposure time than darker regions. Current dithering scripts for MEC enforce a rectangular dithering pattern leading to a non-uniform footprint, but future work will optimize this pattern for maximal uniform coverage.  \label{fig:drizzler}}
\end{figure*}

A common observing strategy with MKID instruments is to dither using a tip/tilt mirror to fill in regions of dead pixels and increase the field of view. A mosaic from these dithered observations may be formed into a spectro-temporal FITS cube by combining each frame onto a common on-sky grid using the \texttt{DrizzlePac} implementation \citep{gonzaga2012drizzlepac} of the \texttt{Drizzle} algorithm \citep{Fruchter2002}. Each frame is mapped onto a sub-sampled output image to generate a single combined image, a spectral cube, a temporal cube, or a spectro-temporal cube with arbitrary wavelength and temporal axes. This allows for the generation of contiguous outputs even with pixel yields of $\sim$ 75\% on active feed-lines \citep{walter2020}, see Figure \ref{fig:drizzler}. 

As all presently supported MKID instruments operate without an image derotator the sky rotation is generally removed from each frame, resulting in an output where every frame is North aligned. An angular differential imaging mode is offered to facilitate interfacing with the Vortex Image Processing package for high-contrast direct imaging (VIP; \cite{2017VIP}) in which each frame in the sequence is aligned so that the first frame is North aligned but the parallactic angle rotation between frames is preserved. 

\subsubsection{Movies}
Movies may be output in GIF or MPEG-4 format and come in two types. The first shows subsequent frames with the desired temporal resolution and run time and is well suited to show rapidly changing features, such as  diffracted speckle patterns that vary on millisecond timescales \citep{2018goebel}. The second format integrates the series of frames and is helpful to illustrate how increasing exposure time affects the final output image. 

\section{\texttt{mkidpipeline}: The MKID Pipeline  Package}\label{sec:implementation}

The MKID Pipeline is implemented as the Python 3 package \texttt{mkidpipeline}\footnote{https://github.com/MazinLab/MKIDPipeline} that includes a corresponding conda environment definition file. The package provides a command-line program, \texttt{mkidpipe}, to process observational data and is configured via three YAML files: \texttt{pipe.yaml}, for general and step specific settings; \texttt{data.yaml}, which defines the data; and \texttt{out.yaml}, which specifies output products. Instructions for basic pipeline setup and execution of a sample dataset are provided in the package \texttt{README}. Complex data processing is expected to require direct use of pipeline methods in a user script. The following subsections describe the pipeline implementation. Additional details may be found in the source code.

\mkp is composed of the modules \texttt{pipeline}, \texttt{photontable}, \texttt{definitions}, \texttt{config}, and \texttt{samples}, along with the sub-packages \texttt{steps}, \texttt{utils}, \texttt{data}, and \texttt{legacy}. Example data and default configuration files are stored in the \texttt{data} and \texttt{config} directories, respectively. 





\subsection{Concept}\label{sec:concept}

The pipeline steps as outlined in \S \ref{sec:dataprocessing} are implemented as modules in \texttt{steps}, with the requirement that each define a \texttt{FlagSet} at \texttt{FLAGS}, a \texttt{StepConfig} (see \S \ref{sec:initandconfig}), and an \texttt{apply()} method. Steps may also implement \texttt{fetch()} when there is a need to compute a persistent calibration data product (CDP), e.g. a wavelength or flat-field calibration solution file. If implemented, \texttt{fetch()} will be provided a path that is guaranteed to be unique for the input data and step configuration used to generate the CDP. This allows multiple users to use these files from a shared location without duplication of effort. 


\begin{sidewaysfigure*}
\vspace{6cm}
\includegraphics[scale=0.5]{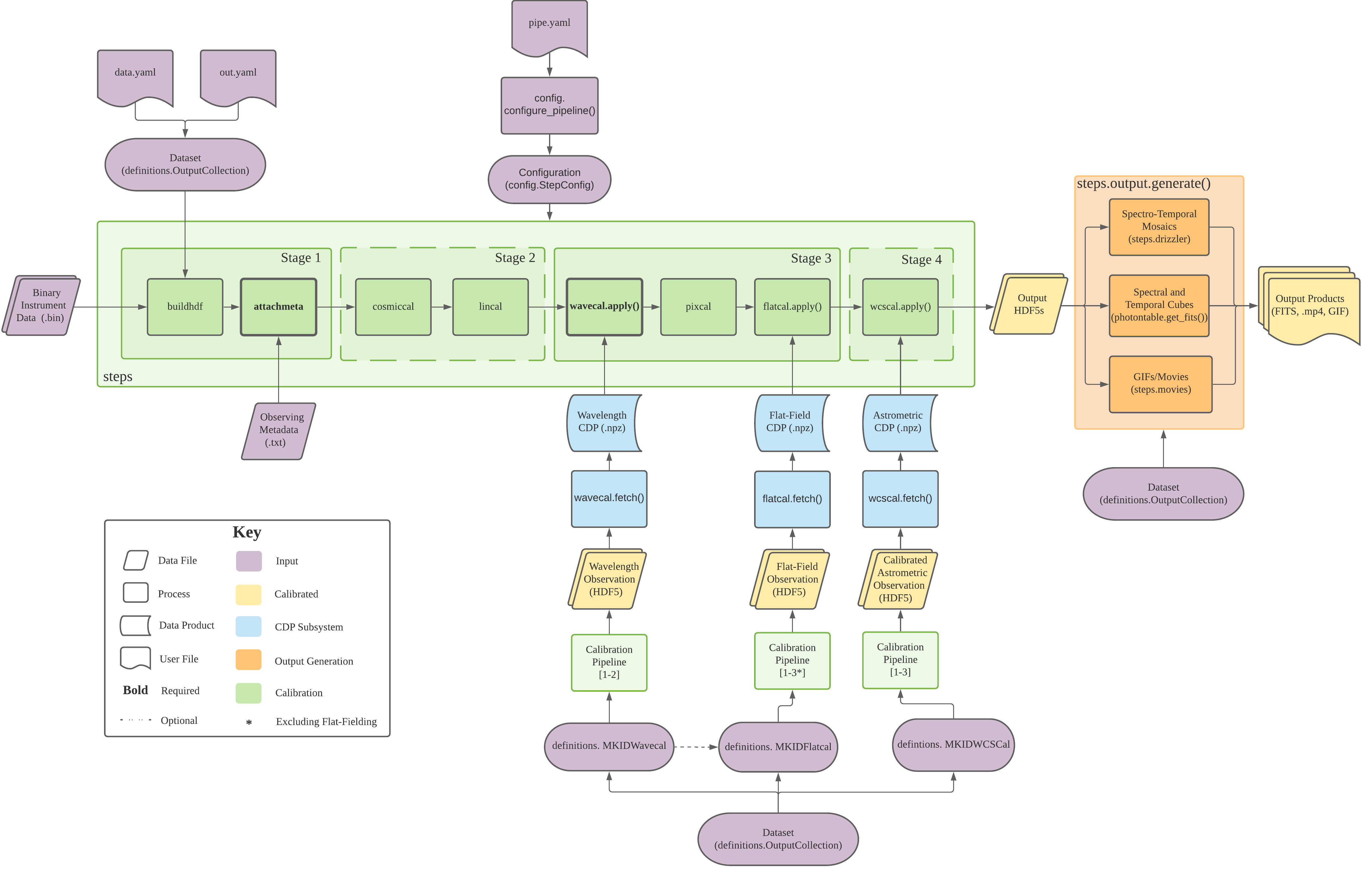}
\label{fig:block3}
\caption{Block diagram of the MKID Pipeline depicting implementation details. 
}  
\end{sidewaysfigure*}

\subsection{Initialization and Configuration}\label{sec:initandconfig}


Each step module with settings is required to implement a subclass of \texttt{config.BaseStepConfig} named \texttt{StepConfig}. In its simplest form, this merely consists of a class-member listing of setting names, default values, descriptions and a YAML tag, though support is provided for additional verification of parameters that may have complex inter-dependencies or depend on other settings from other steps. 

The pipeline places a configuration object for programmatic and interactive use at  \texttt{config.config} after initial configuration (e.g. by loading a \texttt{pipe.yaml}). Access to a fully populated, isolated configuration object is available via the \texttt{PipelineConfigFactory}. This allows individual steps to not worry about whether or not the pipeline has been configured via a file and ensures that required step defaults are present. It also means that any accidental mutations of the configuration do not propagate to other steps or processes.  The configuration object supports parameter inheritance, however default values for individual steps can result in unexpected behaviour as the existence of a child default will take precedence over an explicitly set parent setting. 

When imported, the \texttt{pipeline} module loads steps from the \texttt{steps} sub-package, registering any defined configuration classes with the pipeline YAML loader. These are then available to build a \texttt{config.PipeConfig} of top-level and step specific settings via \texttt{pipeline.generate\_default\_config()} or by loading a config file \texttt{config.configure\_pipeline()}.

In addition to configuration options, the pipeline maintains a set of named flags that may be associated with individual pixels. Flag support is achieved by requiring steps to list any flags they would like to set as a tuple of strings named \texttt{FLAGS}. These are parsed when the steps are loaded and used to build a \texttt{FlagSet} object at \texttt{config.PIPELINE\_FLAGS} that is capable of converting between flag names and bitmasks. The \texttt{FlagSet} is implemented in such a manner to ensure forward compatibility with pipeline data as new flags are added. 

\texttt{mkidpipeline.samples} provides sample data sets and outputs for both programmatic reference and use by the pipeline to generate default \texttt{data.yaml} and \texttt{out.yaml} configuration files during initialization. The resulting files provide comprehensive samples with sensible defaults that may be used to test the pipeline. The raw data is not included due to its extremely large size. 

\subsection{Data Specification}
\label{sec:datasetmanag}
\texttt{definitions} provides classes to manage the description and use of calibration and science data. Data definitions may be created either via class instantiation or via YAML, where support is provided for linking unnested data descriptions by name. For example, an observation may specify a wavecal to use via the name of a top-level wavecal (i.e. not defined explicitly within a different observation) within the same \texttt{data.yaml}. Though possible, it is not generally advised to nest definitions. The \texttt{MKIDObservingDataset} is used to represent  collections of data definitions and defines properties to access key groupings of data: \texttt{<stepname>able} (e.g. \texttt{wavecalable}) are definitions that can have the step applied and  \texttt{<thing>s} (e.g \texttt{wavecals}) are definitions of that thing. 

All observational data are sub-classes or collections of \texttt{MKIDTimerange} objects. This object is defined by a name, a UTC start time (as a Unix timestamp), a stop time or duration, an optional nested \texttt{MKIDTimerange} for a dark exposure, and an optional set of header key overrides. It provides support for metadata retrieval from instrument logs, accessing the associated detector beammap, HDF5 path, and convenience methods for accessing the table of photon data (c.f. \S \ref{sec:photontable}).

Scientific observations are instances of \texttt{MKIDObservation}, which requires the specification of a \texttt{wavecal}, \texttt{flatcal}, and \texttt{wcscal}. Dithered observations are represented by \texttt{MKIDDither} which has similar calibration requirements to \texttt{MKIDObservation}. The dither, however, takes a single \texttt{data} specification which may be either a list of \texttt{MKIDObservation}s, a timestamp within a dither log, or the fully qualified path to a dither log. In the latter two cases, the list of \texttt{MKIDObservation}s is built from the dither, specified calibrations, and any extra header information.

All calibration data sets (including \texttt{MKIDWavecal}, \texttt{MKIDFlatcal}, and \texttt{MKIDWCSCal}) include the \texttt{CalibMixin} mix-in. This provides support for accessing the input time ranges as well as the creation of unique hash strings to identify calibrations made with specific data and settings. Wavelength calibration data sets are represented by an \texttt{MKIDWavecal} and take a list of \texttt{MKIDTimerange}s named by laser wavelength (e.g. `1000 nm') as its data. Flat-field calibrations (\texttt{MKIDFlatcal}) take either a list of \texttt{MKIDObservation}s or the \texttt{name} of an \texttt{MKIDWavecal} as data input. If a \texttt{MKIDWavecal} name is provided then a \texttt{wavecal\_duration} and \texttt{wavecal\_offset} must be given. These specify the duration and starting offset relative to the wavecal's photon tables are used to create new, wavelength calibrated tables for the flat-field calibration.

Astrometric calibration data is represented by an \texttt{MKIDWCSCal} and takes either a platescale, a, \texttt{MKIDDither}, or an \texttt{MKIDObservation} as data. A \texttt{pixel\_ref} and \texttt{conex\_ref} are also required that define a tip/tilt mirror home position and a corresponding pixel location, if applicable. If a dither or observation is used, \texttt{source\_locs} must list the sky coordinates of the targets.

\subsection{Output Specification}
Individual outputs are defined by a named \texttt{MKIDOutput}. This class is defined by a \texttt{name}, a \texttt{data} string specifying a \texttt{MKIDDither} or \texttt{MKIDObservation}, and a \texttt{kind} which specifies the type of output (e.g. movie, drizzle). Optional keys include minimum and maximum wavelength bounds (\texttt{min\_wave}, \texttt{max\_wave}), \texttt{exclude\_flags}, a \texttt{duration}, a \texttt{filename} which specifies the name of the output file, \texttt{units} (photons or photons/s), \texttt{use\_weights} which weights photons by their pipeline weights, \texttt{adi\_mode} which preserves parallactic rotation between drizzled frames (c.f. \S \ref{sec:drizzler}), a \texttt{timestep} which will yield a temporal cube if non-zero, a \texttt{wavestep} which will yield a spectral cube if non-zero, and fields that determine which calibration steps will be applied to the output (e.g. \texttt{wavecal}). If a movie is requested, a \texttt{movie\_runtime} is also required. \texttt{MKIDOutput} provides the pipeline with the properties  \texttt{wants\_<outputtype>} and \texttt{output\_settings} to help determine what output types are needed and what settings need to be used with \texttt{output.generate()} (c.f. \S \ref{sec:output_gen}).

The \texttt{MKIDOOutputCollection} manages the outputs and defines relevant properties to be used by \texttt{outputs.generate()}. These include \texttt{to\_<stepname>} (e.g. \texttt{to\_wavecal}) which gather all of the data definitions needing a particular step given the current configuration, data, and outputs requested. It also provides properties similar \texttt{MKIDObservingDataset} that filter a potentially large data set down to the subset needed for a particular set of outputs.

\subsection{Execution}
\label{sec:output_gen}
The command-line program \texttt{mkidpipe} provides arguments for help, initialization, input verification, and pipeline execution. On initialization, it creates a commented set of pipeline YAML files in the working directory populated with all available settings and a set of default data and output definitions. Re-invoking \texttt{mkidpipe} will validate the files and begin the data reduction process.

On execution, \texttt{mkidpipe} configures the pipeline via \texttt{config.configure\_pipeline()} and then loads the data and output YAMLs by instantiating an \texttt{MKIDOutputCollection}. The data set and outputs are then validated via \texttt{outputs.validation\_summary} and any issues presented to the user for correction. The program then proceeds to call first \texttt{fetch()} and then \texttt{apply()} for each applicable step required for each output. This can be seen diagrammatically in Figure \ref{fig:block3}. Finally, the entire \texttt{MKIDOutputCollection} is fed to  \texttt{outputs.generate()}. This function executes \texttt{photontable.Photontable.get\_fits()} for a spectral or temporal FITS cube, \texttt{movies.fetch()} for a GIF or MPEG-4 output, or \texttt{drizzler.form()} for a combined spectro-temporal mosaic FITS cube as needed. Existing outputs are not, by default, overwritten. 

\subsection{Core Modules and Libraries}
\label{sec:photontable}
Much of this functionality is mediated by the \texttt{Photontable} class described below. The pipeline also depends on AstroPy \citep{astropy:2018}, PyTables \citep{pytables}, and the python2/3 compatible library \texttt{mkidcore}\footnote{https://github.com/MazinLab/MKIDCore}. \texttt{mkidcore} is used for tasks such as logging, flagging, parsing instrument readout information, and managing instrument specific settings. This package ensures compatibility may be maintained across a number of instrument readout systems without editing the pipeline.

The \texttt{photontable} module implements \texttt{Photontable} which handles all interaction with underlying photon data, loading data from and  manipulating the underlying HDF5 file representation. Key functionality is provided to (un)flag pixels, interact with observing metadata, select subsets of photons by wavelength range, time range, and pixel, and form FITS images and cubes (with associated WCS information, if available). Functionality is generally dependent upon what pipeline processing has been completed.

\subsection{Interactive Use}
Users are able to import the \texttt{mkidpipeline} package, create data and output definitions programmatically in a similar manner to that done for pipeline initialization in \texttt{mkidpipeline.samples}. Step operations and numerous utility functions are then available to be used interactively on the data from a terminal.

\section{Summary}

The MKID Pipeline is an open-source extensible pipeline for the reduction and calibration of MKID data. It takes binary per-pixel time-series of photon-induced phase shifts as its input and can perform cosmic ray rejection, linearity calibration, wavelength calibration, flat-fielding, bad pixel masking, and astrometric calibration. This results in calibrated spectro-temporal FITS cubes which can be integrated with traditional astronomical tool chains for scientific analysis. Additionally, unique MKID specific data products, such as time tagged photon lists, can be easily accessed and manipulated for the use and development of new post-processing techniques that utilize photon arrival time statistics. 

The pipeline is designed with automation in mind to allow users to run basic reductions from the command line with unique reductions requiring only the editing of a few configuration files. It also allows future developers to add new algorithms and calibration steps in a modular framework to serve as a base for future MKID instruments and mixed instrument reductions.

\section{Acknowledgements}
S.S. is supported by a grant from the Heising-Simons Foundation. N.Z. was supported throughout this work by a NASA Space Technology Research Fellowship. K.K.D. is supported by the NSF Astronomy and Astrophysics Postdoctoral Fellowship program under award \# 1801983. The authors would like to thank Clarissa Rizzo, Joshua Breckenridge, and Xiaofei Zhang who helped with  testing at various stages of pipeline development and improved the quality of this work.

\bibliography{pipeline}{}
\bibliographystyle{aasjournal}



\end{document}